\documentclass[conference,onecolumn]{IEEEtran}
\IEEEoverridecommandlockouts

\usepackage{cite}
\usepackage{amsmath,amssymb,amsfonts}
\usepackage{algorithmic}
\usepackage{graphicx}
\usepackage{textcomp}
\usepackage{xcolor}
\usepackage{booktabs}
\def\BibTeX{{\rm B\kern-.05em{\sc i\kern-.025em b}\kern-.08em
    T\kern-.1667em\lower.7ex\hbox{E}\kern-.125emX}}
    
\begin{document}

\title{Fast Session Resumption in DTLS\\ for Mobile Communications\footnotemark
}

\author{\IEEEauthorblockN{Gyordan Caminati, Sara Kiade, Gabriele D'Angelo, Stefano Ferretti, Vittorio Ghini}
\IEEEauthorblockA{Department of Computer Science and Engineering\\
University of Bologna\\
Bologna, Italy\\
\{gyordan.caminati, sara.kiade\}@studio.unibo.it 
\{g.dangelo, s.ferretti, vittorio.ghini\}@unibo.it}
}

\maketitle

\footnotetext{
The publisher version of this paper is available at \url{https://doi.org/10.1109/CCNC46108.2020.9045119}.
\textbf{{\color{red} This is the pre-peer reviewed version of the article: ``Gyordan Caminati, Sara Kiade, Gabriele D'Angelo, Stefano Ferretti, Vittorio Ghini. Fast Session Resumption in DTLS for Mobile Communications. Proceedings of the IEEE Consumer Communications and Networking Conference 2020 (CCNC 2020).''.}}}

\begin{abstract}
DTLS is a protocol that provides security guarantees to Internet communications. It can operate on top of both TCP and UDP transport protocols. Thus, it is particularly suited for peer-to-peer and distributed multimedia applications. The same holds if the endpoints are mobile devices. In this scenario, mechanisms are needed to surmount possible network disconnections, often arising due to the mobility or the scarce resources of devices, that can jeopardize the quality of the communications. Session resumption is thus a main issue to deal with. To this aim, we propose a fast reconnection scheme that employs non-connected sockets to quickly resume DTLS communication sessions. The proposed scheme is assessed in a performance evaluation that confirms its viability.
\end{abstract}


\section{Introduction}
DTLS (Datagram Transport Layer Security) is a secure communication protocol, often utilized in peer-to-peer and multimedia distributed applications. For instance, DTLS is at the basis of WebRTC technologies, that promote the development of in-browser applications such as audio and video conferences, chats, file transfer and online gaming~\cite{Desprat:2015}. In these contexts, enabling end-points to exchange data often requires authentication of the involved parties. Not only, integrity and confidentiality of exchanged data must also be ensured~\cite{rfc7925}. DTLS is thought for these purposes. In particular, DTLS is aimed at providing secure encryption and authentication services for UDP. Compared to the Transport Layer Security (TLS), which has been devised to work over TCP, the DTLS protocol differs in several areas due to the unreliability of UDP~\cite{WeiD18}. In fact, DTLS has to deal with the unreliability and absence of ordering delivery guarantees of the underlying network stack.

DTLS can be of great help even when the interacting end-points are mobile, or even in case of IoT devices~\cite{gda-hpcs-16,gda-simpat-iot,Moosavi2018432,Rajagopalan2017,gda-concurrency-iot}. In fact, designing a trustworthy networking framework is essential for secure mobile and pervasive services~\cite{7777209}. And needless to say, in this application domain, dealing with security aspects is even worsened considering the constrained resources of the involved devices~\cite{Cho:2019}.

As concerns multimedia, distributed and peer-to-peer applications, such as online gaming, multimedia live streaming and so on, an important feature to guarantee is the maintenance of the communication session between two (or multiple) end-points. In fact, the notion of ``session'' allows the end-points to restart a communication, even after a network disconnection occurred. A session resumption mechanism should allow restarting the interactions from the exact point where the two parts of the communication process arrived, thus avoiding the need to restart from scratch~\cite{Aviram2019117}.

There are different approaches to resume a session in a DTLS-based communication. Session IDs (i.e.~identifiers) can be employed to identify and resume a previously established session. In this case, both end-points should maintain this ID. Alternatively, in order to relieve the server from maintaining the session IDs related to all the connected clients, a session ticket can be utilized. This session ticket is stored on the client side. In case of disconnections, the client can employ such ticket when reconnecting to the server. Finally, it is possible to devise protocol extensions to identify the session. A common problem to all these approaches is due to the overhead needed to resume a session, since the previously established connections must be closed by both parties and a (shortened) handshake must be performed to resume the session, in anyway.

In this paper, we propose a new a strategy, called \emph{Fast Resumption}, that exploits non-connected sockets at the server side to resume a DTLS session. We implemented two variants of this approach. The first one, Implicit Port Communication (\emph{IPC}), consists in employing at the server side a non-connected socket, bound to a port which is different from the one employed for listening to novel connections. Through the use of specific cookies (i.e.~a small amount of information stored in the client), the session can be quickly resumed. However, this approach suffers the limitations caused by the presence of restricted NATs (Network Address Translation) gateways. In fact, if the client changes its IP address during a handover then the NAT might filter messages sent from an unknown client to a port different from that specified in the WelcomeSocket (i.e.~the socket used in the server to accept new connections).

The second implementation is called Temporary Connected Socket (\emph{TPC}). In this case, a temporary socket is used to create the session. Simply put, once a client wants to establish the session, it contacts the server to the port where it is listening for novel connections. During the handshake, a temporary socket is created and then used to let the server and the client establish a novel non-connected socket. While a slightly higher overhead is required during the first handshake, this solution allows surmounting the presence of NATs. At the same time, it enables the client to change its IP address, perform handovers and reconnect to the server in a seamless way.

We performed a testbed evaluation to assess the performance of Fast Resumption when compared to the traditional DTLS communications. The results demonstrate that Fast Resumption is able to speed up the communications with respect to standard DTLS.\\

The remainder of this paper is organized as follows. Section~\ref{sec:back} provides the background needed on DTLS and session resumption schemes. Section~\ref{sec:fr} describes the proposed scheme for fast resumption in DTLS based communications. In Section~\ref{sec:perf} we provide results from an experimental evaluation we conducted to assess the viability of the approach. Finally, Section~\ref{sec:conc} provides some concluding remarks.

\section{Background}\label{sec:back}

\subsection{Datagram Transport Layer Security}
Datagram Transport Layer Security (DTLS) is a protocol that provides security for datagram-based applications. It is widely used in multimedia distributed applications and it is at the basis of the WebRTC technology, which is used in many browser-based multimedia applications~\cite{Desprat:2015}. Derived from (and similar to) the Transport Layer Security (TLS) protocol, DTLS is a stream oriented protocol that offers security features to prevent eavesdropping, tampering, or message forgery. However, the basic design philosophy of DTLS is to construct TLS over unreliable datagram transport~\cite{rfc6347}. Thus, DTLS is built on top of UDP rather than TCP. The protocol explicitly copes with the absence of reliability and ordered delivery guarantees, that are not provided by UDP. In particular, with respect to TLS, DTLS adds explicit sequence numbers to the messages and retransmission timers that are used during the session handshake.
   
A typical communication scenario via DTLS is as follows. For the sake of a simpler description, let denote the end-point that begins the communication as the client, and the end-point that receives the communication request as the server. As a matter of fact, the approach works either if the distributed application is based on a client-server distributed architecture or if a peer-to-peer architecture is employed. A server listens for novel clients on a specific address IPAddr:Port. This address is used on multiple active sockets. Each communication channel with an active client uses a connected socket that is bound on a specific port. In this scheme, the kernel is in charge of dispatching every message arriving to that address, based on the connected socket, that identifies the communication channel. Moreover, a single non-connected socket, hereinafter called the \emph{WelcomeSocket}, is exploited to handle novel clients. Since the socket is not connected, each message from a novel client sent to that port is dispatched by the kernel to the WelcomeSocket. At the client side, the connected socket is used to interact with the server. Messages arriving to the port associated with the socket, but which do not come from the server and from the specific socket, are discarded. This approach works pretty well if the end-points are stable and able to maintain fixed IP addresses during the whole life of the communication channel. In fact, a problem arises if, for some reason, an end-point (let's say the client) performs a handover with a change of address. In fact, if the client changes IP address and sends a new message to the server, this message will be interpreted as coming from a novel host (i.e.~the communication channel with the server is still bound to the old IP address). Thus, the message would be dispatched to the WelcomeSocket. At this point, the process listening on the WelcomeSocket probably would be unable to recognize the message, since it expects to receive a new initialization message to perform a handshake.

\subsection{Session Resumption}
Session resumption is a feature of core TLS/DTLS specifications that allows a client to continue with an earlier established session state, in case of disconnections~\cite{Moosavi2018432,Rajagopalan2017,rfc7925,SyBFF18}. It is a vital feature to guarantee, especially when dealing with mobile devices, that can experience connection losses and horizontal/vertical handovers~\cite{Ferretti2016390,Ferretti2013481}. In the rest of this section, we will describe some possible solutions which have been devised to support session resumption. It is worth mentioning that these approaches are usually thought to work in client-server architectures.

\subsubsection{Session Resumption through Session ID}
A method to resume a session is to exploit session IDs. Both parties (i.e.~client and server) store session details in their cache. Once the client tries to reconnect after a disconnection, it sends a \emph{ClientHello} message that includes the previously determined session ID. Upon reception of this message, the server can decide if that session can be resumed. If this is the case then the client and server perform a shortened handshake.

\subsubsection{Session Resumption through Session Ticket}
This proposal is specifically though for client-server distributed architectures. The scheme is cited in DTLS related RFCs, even if we are not aware of specific available implementations. In particular, in \cite{rfc8094} it is claimed that ``it is highly advantageous to avoid server-side DTLS state and reduce the number of new DTLS sessions on the server that can be done with TLS Session Resumption without server state''. And then in \cite{rfc7925} it is stated that it is possible to devise a ``TLS/DTLS session resumption that does not require per-session state information to be maintained by the constrained server. This is
accomplished by using a ticket-based approach''. Session tickets are encrypted data, which are created by the server, that contain the information related to a session. This data can be used once the session has to be resumed. Thus, session resumption through session tickets can be implemented to shorten the resumption handshake, with all parameters that are specified by the client. This relieves the server to maintain a database with information on all active communications.

\subsubsection{Session Resumption through Connection ID}
This solutions is a proposed draft, without a reference implementation being provided~\cite{ietf-tls-dtls-connection-id-06}. This mechanism aims at solving the problems concerned with the possibility that a client node changes its IP address while trying to resume a previously established session. In essence, the idea is to extend the protocol adding a data field (i.e.~the ConnectionId) to the ClientHello and ServerHello messages, exchanged at the beginning of the DTLS session only. The two endpoints can retrieve the session through this data field.

\subsubsection{Limitations of Session Resumption}
Whatever the specific mechanism being used, session resumption has some inherent limitations. First, the previous connection and session has to be closed by both parties, before resuming. This leads to some delays. Second, a handshake is always needed, even if it is shortened in the number of exchanged messages. Again, this leads to some delays. Third, someone (client, server or both) has to maintain some information on the session, in order to resume it.

\section{Fast Resumption}\label{sec:fr}
In this work we present our novel proposal called \emph{Fast Resumption} that is based on the idea to exploit non-connected sockets at the server side. Each communication with a different client is then performed using a non-connected socket, bound to a different port.

In order to better explain the rationale behind this solution, let consider a use case example, based on the typical scenario of a server that binds more sockets to the same address. Let's  assume that we have a client that uses a UDP socket (ClientSocket) bound to the address 184.16.1.30:1234. The server, on the other hand, maintains two sockets bound to the same IP-address:Port 108.110.11.12:4433.
The two sockets are:
\begin{enumerate}
    \item WelcomeSocket: this socket is used to establish novel communications with novel clients. Every novel client can start a handshake with the server sending a message to this socket. Since the server receives messages from multiple hosts through this socket, the welcome socket is non-connected.
    \item ComSocket: this is a socket used to communicate with the client. Hence, this socket is connected to ClientSocket.
\end{enumerate}

Since at the server both sockets work on the same address, 108.110.11.12:4433, the kernel looks at the address of each message received on that port. If the sender is ``ClientSocket'' then the message is dispatched to ComSocket, otherwise the message is dispatched to WelcomeSocket. Thus, if the client disconnects and then tries to resume a communication using a different IP address, due to a handover, its messages will be dispatched by the kernel to the WelcomeSocket. Once received, the process working on WelcomeSocket will treat the message as part of a novel connection. Hence, the message will be discarded since it will not respect the handshake protocol (in this scenario, the client is trying to continue its communication with the server, without performing a reconnection from scratch, while on WelcomeSocket the server is listening for novel ``hello'' messages). We might decide to assign a novel port to ComSocket. But if ComSocket is connected then the messages sent by the client after the handover (hence, with a novel address) will be discarded. Vice versa, if ComSocket is bound to an address, which is different from WelcomeSocket, and ComSocket is not connected, then messages coming from the novel client address will be delivered to the server application. However, this approach has to cope with the possible presence of NAT gateways, as discussed in the rest of this section.

\subsection{IPC (Implicit Port Communication)}
\begin{figure*}[tb]{
\centering
\includegraphics[width=1.00\textwidth]{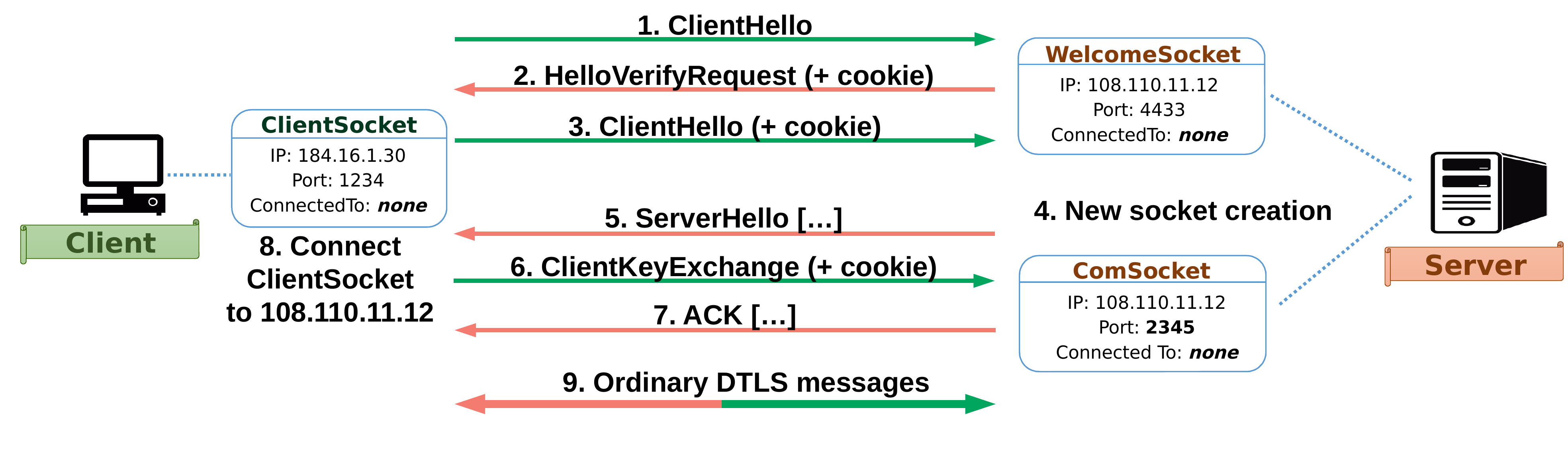}
\caption{Fast resumption -- IPC: message exchange.\label{fig:ipc}}
}
\end{figure*}

According to this scheme, upon disconnection the communication is resumed by using a different non-connected socket. This is a lightweight approach, that does not require any modification to the DTLS protocol. The handshake is performed as follows (see Figure~\ref{fig:ipc}):
\begin{enumerate}
    \item The client contacts the server through the WelcomeSocket (i.e.~108.110.11.12:4433);
    \item The server answers with a \emph{HelloVerifyRequest} message, that contains a cookie;
    \item The client sends a \emph{ClientHello} message with the received cookie;
    \item the server creates a novel, non-connected socket, bound to a different port (e.g.~2345);
    \item the server uses such novel socket to send a \emph{ServerHello} to the client;
    \item the client answers the message with an acknowledgment;
    \item the server concludes the handshake;
    \item at the client side, the ClientSocket is connected to ComSocket;
    \item the DTLS communication channel is established and ordinary DTLS messages can be exchanged among the two endpoints.
\end{enumerate}

The created communication channel is resistant to IP changes and reconnections, since both the client and the server maintain a shared cookie that can be used to identify the previously established channel and then to resume the communication.

\begin{figure*}[tb]{
\centering
\includegraphics[width=1.00\textwidth]{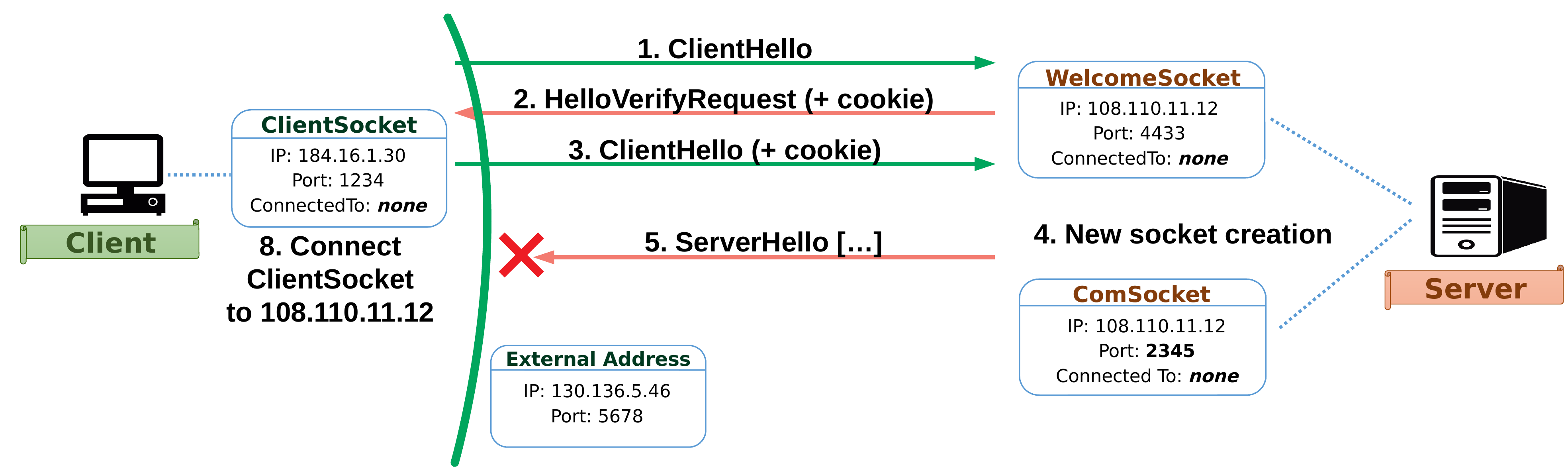}
\caption{Fast resumption -- IPC: problems due to the presence of a NAT gateway.\label{fig:nat}}
}
\end{figure*}

\subsubsection{Pros}
This approach does not require any additional resource. A traditional handshake is exploited and no novel messages are required between the two endpoints. If one endpoint changes its IP address then the DTLS communication channel is not broken, thank to the generated cookie.

\subsubsection{Cons}
If the client is placed behind a port-restricted NAT (or more restricted NAT versions), the ServerHello message from the sever is discarded by the NAT policies, since it comes from an unknown address (see Figure \ref{fig:nat}). This limitation is quite relevant since nowadays most clients are placed behind NAT gateways.

\subsection{TCS (Temporary Connected Socket)}
\begin{figure*}[tb]{
\centering
\includegraphics[width=1.00\textwidth]{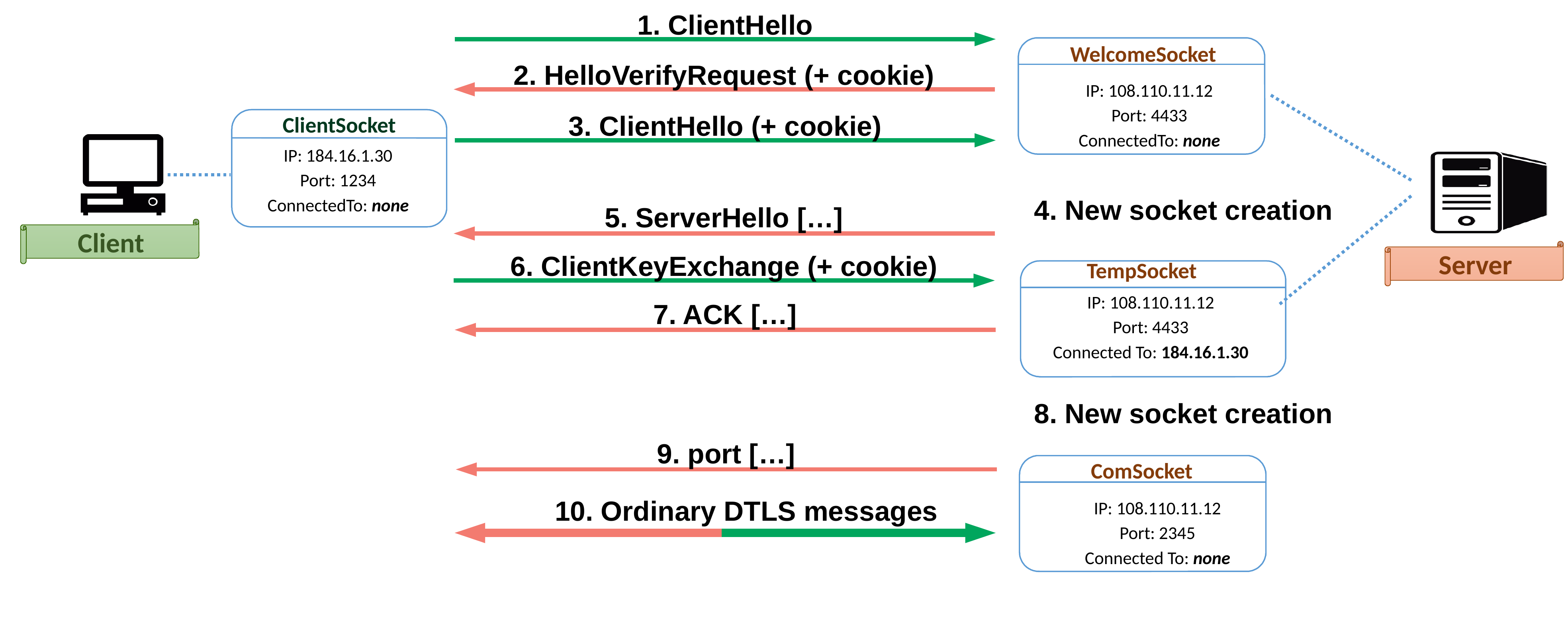}
\caption{Fast resumption -- TCS: message exchange.\label{fig:tcs}}
}
\end{figure*}

An alternative approach is to exploit a temporary socket, bound to the same port of the WelcomeSocket and connected to the client. This temporary socket lasts the time required to perform the first handshake, only. During the handshake, the server creates a different, non-connected socket that is bound to a different port. This new socket will be used for the rest of the communication. More in detail, the scheme works as follows (see also Figure~\ref{fig:tcs}):

\begin{enumerate}
    \item The client sends a \emph{ClientHello} message to the server WelcomeSocket 108.110.11.12:4433;
    \item the server answers with a \emph{HelloVerifyRequest}; this message  contains a cookie;
    \item the client sends another \emph{ClientHello} message embodying the received cookie;
    \item the server creates a novel socket, bound to the same port of the WelcomeSocket (4433), which is connected to the client address (184.16.1.30:1234);
    \item the server sends a \emph{ServerHello} message to the client from the novel socket, hence freeing the WelcomeSocket. This strategy surmounts problems concerned to the presence of a NAT, since the novel socket is bound to the same port of the WelcomeSocket, i.e.~an address recognized by the NAT;
    \item the client continues the handshake sending messages to the same address (184.16.1.30:1234). On the server side, messages are dispatched by the kernel to the novel socket (connected to the client);
    \item the server concludes the handshake on this temporary novel socket, by sending an ACK message;
    \item the server creates a novel socket (non-connected and bound to a different port) to be used with the client after the handshake. This novel socket is needed, since the temporary socket is connected to the actual client address. Hence, if the client would change its address, due to a handover, the messages sent from this novel address would be dispatched, on the server side, to the WelcomeSocket, that would discard them, waiting for a \emph{ClientHello} message (as already discussed);
    \item the server sends to the client the novel address that it must use to send messages hereinafter;
    \item now a connection is established and messages can be sent using DTLS.
\end{enumerate}

Since the final socket used for the communication is non-connected, the client is able to perform handovers and change its address during the communication. The security of the communication is guaranteed by the combined use of the DTLS security features and the cookie usage, that provides the authentication of both end-parties.

\subsubsection{Pros}
This approach surmounts the presence of restricted NATs. A novel socket is created only once the client identity has been verified by the sender. This protects the scheme from IP spoofing attacks. No modifications on the protocol are required.

\subsubsection{Cons}
This solution requires more messages to be exchanged between the end-points with respect to IPC. Two sockets are created to initiate the communication, i.e.~a temporary socket is exploited to establish another one, before being destroyed. However, this sockets lasts only during the initial handshake. Moreover, it is necessary to handle the possible loss of the message, sent by the server to the client, containing the address to be used. In our current implementation, a timeout is set, after which the sender sends the message again, until the sender receives a message from the client to the novel socket.

\section{Performance Evaluation}\label{sec:perf}
In order to assess the performance of the Fast Resumption mechanism, we have setup a testbed composed of two Virtual Machines (VMs) running on the same host. Each VM is equipped with GNU/Linux Ubuntu 18.04 LTS while the host machine is with Windows 10. The first machine (\emph{VM-C}) runs a client application written in C language that implements a DTLS client that exchanges messages with a server application running on \emph{VM-S}. The client opens a DTLS socket, connects to the server and then sends messages to the server that replies with acknowledgments. To test the Fast Resumption mechanism under realistic assumptions, the VM-C has been modified to add controlled network delays and to simulate handovers, that shutdown the network interface used by the client to communicate.

Due to space limitations, in this paper we report only the performance evaluation of the TCS variant of Fast Resumption (DTLS-TCS). In practice, the two proposed variants DTLS-TCS and DTLS-IPC differ only in how the session and the encrypted communication channel between the client and the server are established. In the case of the DTLS-TCS variant, a slightly higher overhead is required during the first handshake, but this solution allows surmounting the presence of NATs. After the session has been established, both variants allow the client to change his IP address, perform handovers and reconnect to the server in a seamless way. Since the DTLS-TCS variant can be used in more network setups than DTLS-IPC and both have the same performance in case of handovers, we preferred to assess the behavior of DTLS-TCS.

We start comparing the performance of DTLS-TCS with the standard DTLS when used by a client with a single network interface and with a delay that is added to every incoming and outgoing network packet. Moreover, every 10 seconds the network interface in the client is shutdown and then reactivated. This leads to the triggering of Fast Resumption in DTLS-TCS and the setup of a brand new connection in the standard DTLS. In this scenario, a new connection setup can be initialized only when both sides of the communication decide that the connection is interrupted. This is usually implemented by means of application-layer timeouts. In this performance evaluation, this timeout has been set to 1 second.

The metric that we have used to compare the protocols is very simple since we measure the amount of time (Wall-Clock Time, WCT) that is necessary by the client to receive 600 acknowledgments that correspond to the same number of messages correctly sent by the client and received by the server. Clearly, if a message (or its acknowledgment) is lost due to a change in the network connectivity then the client has to transmit the message again, leading to an increase of the amount of time to complete this test. In other words, the best protocol is the one that is able to complete the test in less time. Since DTLS-TCS supports network multi-homing, we repeated the previous experiments with a virtual machine equipped with two network interfaces. Also in this case, the client network interfaces are shutdown every 10 seconds (interleaved 5 seconds) to simulate handovers. Even if this setup is quite simplistic, it can provide a preliminary, yet neat, evaluation of the performance offered by DTLS-TCS in a multi-homing environment.

\begin{figure}[tb]{
\centering
\includegraphics[width=1.00\columnwidth]{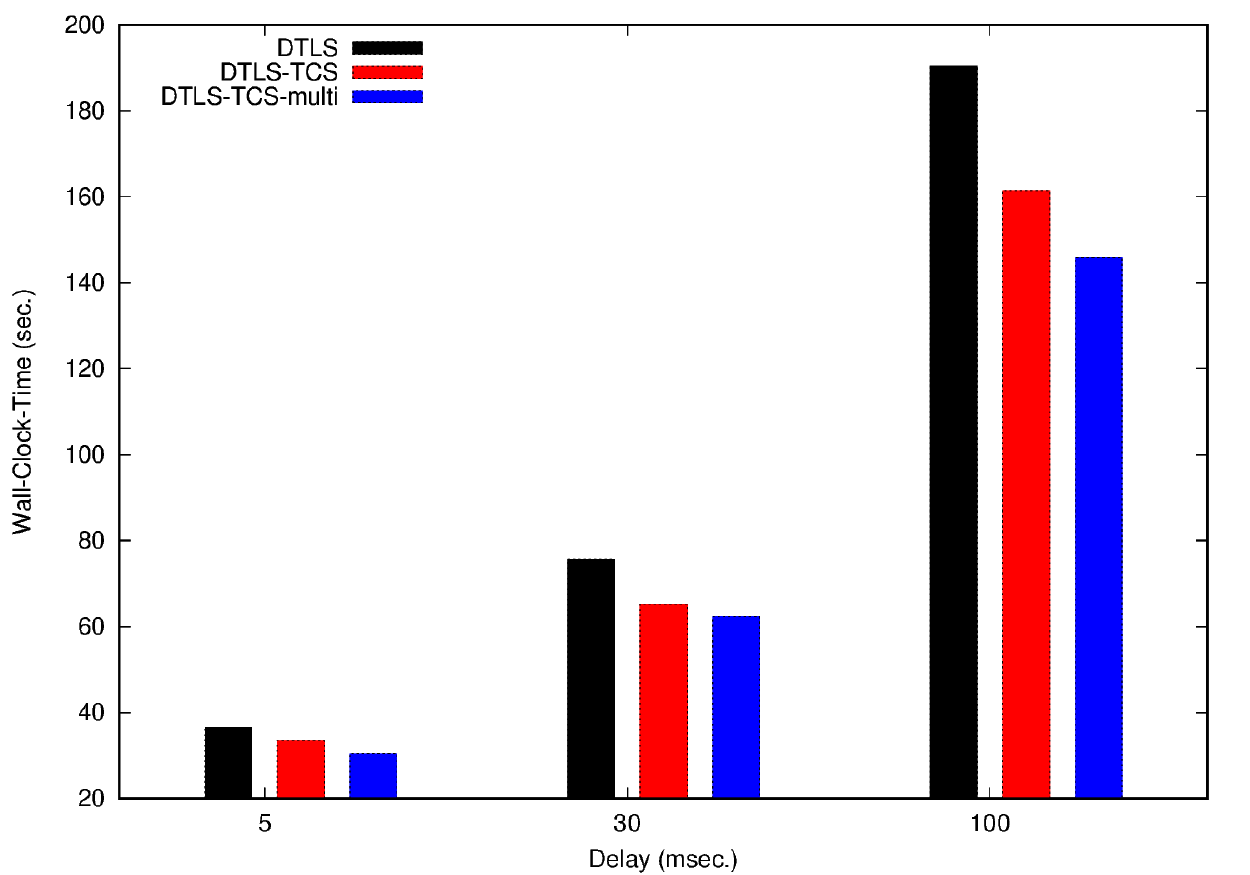}
\caption{Wall-Clock-Time (WCT) comparison of DTLS with DTLS-TCS mono and multi-home versions (lower is better).\label{fig:wct}}
}
\end{figure}

In Figure~\ref{fig:wct} are reported the results obtained averaging multiple runs in the setup described above and considering an increasing amount of delay added to each packet (i.e.~5-100 msec.). As expected the standard version of DTLS is the slowest since the DTLS-TCS version is able to significantly reduce the amount of time required to complete the experiment. Moreover, the multi-home version of DTLS (labelled as DTLS-TCS-multi) is even able to speedup the execution with respect to DTLS-TCS. The gain obtained by both DTLS-TCS and DTLS-TCS-multi increases in presence of a higher delay experienced by the network packets flowing between the client and the server. 

Table~\ref{Table:perc_gain} reports the performance gain (in percentage) of DTLS-TCS and DTLS-TCS-multi with respect to the standard DTLS. The performance gain obtained by DTLS-TCS-multi with respect to DTLS goes from $16.61$\% with delay is 5 msec up to $23.42$\% for 100 msec.

\begin{table}[h]
  \centering
  \caption{Performance gain (in percentage) of the Fast Resumption variants with respect to the standard DTLS.}
\begin{tabular}{ccc}
\toprule
\textbf{Delay (msec.)}   & \textbf{DTLS-TCS} & \textbf{DTLS-TCS-multi} \\
\midrule
\textit{5}      & 8.35\%   &    16.61\% \\
\textit{30}     & 13.63\%  &    17.48\% \\
\textit{100}    & 15.22\%  &    23.42\% \\
\bottomrule
\end{tabular}
\label{Table:perc_gain}
\end{table}

At a later stage, the source code of DTLS-TCS and the testbed applications will be made freely available on the research group homepage~\cite{anansi}.

\section{Conclusions}\label{sec:conc}
In this paper, we proposed an approach, called Fast Resumption, aimed at surmounting network disconnections in mobile communications based on the DTLS protocol. In presence of network handovers, the proposed scheme is able to continue the communication without requiring the setup new DTLS connections. This significantly reduces the amount of time in which the mobile devices are unable to communicate, shortening the time spent in reconfiguring the communication. The saving is even more relevant in presence of mobile nodes equipped with multiple network interfaces since that the Fast Resumption scheme supports network multi-homing. As a future work, we plan to evaluate the proposed mechanism in more complex setups and to study the effect of multi-homing on innovative applications, based for instance on the use of distributed ledgers~\cite{gda-ethereum}.

\bibliographystyle{IEEEtran}
\bibliography{biblio}

\end{document}